\documentclass[conference]{IEEEtran}
\IEEEoverridecommandlockouts
\usepackage{ifpdf}

\usepackage{cite}

\ifCLASSINFOpdf
\else
\fi
\usepackage[cmex10]{amsmath}
\usepackage{amssymb,amsfonts}

\usepackage{algorithm}
\usepackage[noend]{algpseudocode}

\usepackage{array}

\usepackage{mdwmath}
\usepackage{mdwtab}

\usepackage{eqparbox}
\usepackage{fixltx2e}

\usepackage{stfloats}

\usepackage{url}

\usepackage{theorem}

\theoremstyle{plain}			

\newtheorem{thm}{Theorem}[section]

\newtheorem{defn}[thm]{Definition}

\newcommand{\qed}{\hfill \ensuremath{\Box}}
\newcommand{\cat}{~\|~}

\renewcommand{\vector}[1]{\mathbf{#1}}
\renewcommand{\matrix}[1]{\mathbf{#1}}
\newcommand{\rexec}{\stackrel{\$}{\longleftarrow}}
\newcommand{\chiexec}{\stackrel{\chi}{\longleftarrow}}
\newcommand{\checkE}{\stackrel{?}{=}}

\begin{document}
%
\title{LWE-based Identification Schemes}

\author{
\IEEEauthorblockN{Rosemberg Silva\IEEEauthorrefmark{1}}
\IEEEauthorblockA{Institute of Computing\\
University of Campinas\\
S\~{a}o Paulo, Brazil 13083-852\\
Email: rasilva@ic.unicamp.br}
\and
\IEEEauthorblockN{Antonio C. de A. Campello Jr.\IEEEauthorrefmark{2}\thanks{Work supported by the State of S\~{a}o Paulo Research Foundation under grant 2008/07949-8\IEEEauthorrefmark{1}  and 2009/18337-6\IEEEauthorrefmark{2}}
\IEEEauthorblockA{Institute of Mathematics, Statistics \\ and Computer Science \\
University of Campinas \\
S\~{a}o Paulo, Brazil 13083-859 \\
Email: campello@ime.unicamp.br}
}
\and
\IEEEauthorblockN{Ricardo Dahab}
\IEEEauthorblockA{Institute of Computing\\
University of Campinas\\
S\~{a}o Paulo, Brazil 13083-852\\
Email: rdahab@ic.unicamp.br}}


%


\maketitle

\begin{abstract}
Some hard problems from lattices, like LWE (Learning with Errors), are particularly suitable for
application in Cryptography due to the possibility of using worst-case to average-case reductions
as evidence of strong security properties. In this work, we show two LWE-based constructions of zero-knowledge
identification schemes and discuss their performance and security. We also highlight the design choices that make our solution
of both theoretical and practical interest.
\end{abstract}


%
\IEEEpeerreviewmaketitle

\section{Introduction}
\subsection{Identification Schemes}

A zero-knowledge protocol can be employed as a way of demonstrating the knowledge of a secret without actually revealing it. Several instantiations of this idea have been proposed since the seminal work of Fiat and Shamir \cite{fiat86}. The code-based constructions of Stern \cite{Stern94anew} and V\'{e}ron \cite{DBLP:journals/aaecc/Veron96} have recently been adapted to the lattice domain, as seen in the work of Kawachi, Tanaka and Xagawa \cite{kawachi08}, and Cayrel, Lindner, R\"{u}ckert and Silva \cite{DBLP:conf/provsec/CayrelLRS10}. In the present work, we propose alternative lattice-based zero-knowledge constructions, using as security assumption the hardness of the LWE Problem, defined below. As seen with several other lattice hard problems, LWE exhibits worst-case to average-case reduction, which is a very interesting property for cryptographic systems because it increases one's confidence that a random instance of the system is indeed hard to break. 

\subsection{The LWE Problem}
The \textit{Learning with Errors} (LWE) problem was first proposed by Regev in \cite{DBLP:conf/STOC/Regev05} as an extension of the so-called \textit{Learning from Parity with Noise} (LPN) problem. The search version of LWE can be regarded, roughly speaking, as the problem of recovering a secret $s$ given a ``noisy linear equation'' of the form $\matrix{A}\vector{s} \approx \vector{y} \mbox{ mod } q$. More formally, we can state LWE as follows \cite{DBLP:conf/STOC/Regev05}:
\begin{defn}[Learning with Errors]
\label{lwe}
Let $q$ be a prime number and $\chi$ a probability distribution on $\mathbb{Z}_q$. Given a \textit{secret} $\vector{s} \in \mathbb{Z}_q^n$, we denote by $\matrix{A}_{\vector{s},\chi}$ the probability distribution on $\mathbb{Z}_{q}^n \times \mathbb{Z}_q$ obtained by choosing a vector $\vector{a} \in \mathbb{Z}_q^n$ uniformly at random, choosing $\vector{e} \in \mathbb{Z}_q$ according to $\chi$, and outputting $(\vector{a}, \left\langle \vector{a},\vector{s} \right\rangle + \vector{e} \mod q)$. We define (the search version of) LWE$_{q,\chi}$ as the problem of recovering the secret $\vector{s}$ (with high probability) given an arbitrary number of samples from $\matrix{A}_{\vector{s}, \chi}$. The decision version of LWE$_{q,\chi}$ is the problem of distinguishing $\matrix{A}_{\vector{s},\chi}$ samples from the uniform distribution on $\mathbb{Z}_q^n \times \mathbb{Z}_q$. For $q = 2$ this is the well-known LPN problem.
\end{defn}

In recent years, the LWE problem has often been used as the basis of many cryptographic constructions due to its strong security properties such as the decision-to-search and average-to-worst-case reductions. Moreover, if $\chi$ is the ``discrete Gaussian'' distribution with standard deviation $\alpha q > 2/\sqrt{n}$, the hardness of LWE is proven to be related to the worst case of some well known lattice problems such as {\begin{footnotesize}GAP\end{footnotesize}\normalsize{SVP} and SVIP, which are believed to be secure even under quantum attacks. For practical applications, it is worth considering the ``ideal'' version of LWE, or ring-LWE, in which the entries of $\vector{a}$ are the coefficients of polynomials in the ring $\mathbb{Z}_q/ \left\langle x^n+1 \right\rangle $, the ``anticyclic'' ring. This approach allows faster matrix-vector multiplications due to Fast Fourier Transforms and reduced keys of size equal to $O(n)$ elements of $\mathbb{Z}_q$. Also, a reduction from the approximate version of SVP on ideal lattices to the search version of ring-LWE under certain assumptions on $\chi$ (what leads to a non-spherically distributed error) was recently published~\cite{DBLP:conf/eurocrypt/LyubashevskyPR10}. 

It is worth noting that LWE is closely related to the SIS problem, which is that of finding a \textit{small integer solution} $\vector{x}$ for the equation $\matrix{A}\vector{x} = 0 \mbox{ mod } q$ (or $\matrix{A}\vector{x} = \vector{y} \mbox{ mod } q$ for the inhomogeneous SIS). Concerning identification schemes, some of them are built upon SIS and its version on ideal lattices \cite{DBLP:conf/provsec/CayrelLRS10}. An ID-scheme based on a slight modification of LWE is proposed in \cite{DBLP:draft/Xagawa07} but it is not clear whether this modification preserves LWE hardness. Thus, the schemes proposed in this work are, to our knowledge, the first ones with security based on the hardness of LWE.

\subsection{Auxiliary Primitives}

\begin{defn}[Hamming isometry $\Pi_{\gamma, \Sigma}$]  \label{wtprestrans}
Let $\Sigma$ be 
a permutation of $\{1,\ldots,n\}$ and $\vector{\gamma} = ({\gamma}_1, \ldots, {\gamma}_n) \in \mathbb{F}_q^n$ 
such that ${\gamma}_i \neq 0, \forall i$. We define the transformation $\Pi_{\gamma, \Sigma}$ as the mapping $\mathbb{F}_q^n \rightarrow \mathbb{F}_q^n$,
taking $\vector{v}$ to ${\gamma}_{\Sigma(1)}v_{\Sigma(1)}, \ldots,{\gamma}_{\Sigma(n)}v_{\Sigma(n)}$.
\end{defn}

This transformation scales through the multiplication by a scalar and preserves the Hamming weight. That is, 
$\forall \alpha \in \mathbb{F}_q$ and $\forall \vector{v} \in \mathbb{F}_q^n$, we have 
$\Pi_{\vector{\gamma}, \Sigma}(\alpha \vector{v}) = \alpha \Pi_{\vector{\gamma}, \Sigma}(\vector{v})$~and~
$\text{wt}(\Pi_{\vector{\gamma}, \Sigma}(\vector{v})) = \text{wt}(\vector{v})$.

\subsection{Notation and Conventions}
\begin{itemize}
	\item $V_h, ~P_h$ - honest Verifier, honest Prover
	\item $V_c, ~P_c$ - cheating Verifier, cheating Prover
	\item $V^{\prime}, ~P^{\prime}$ - arbitrary Verifier, arbitrary  Prover
	\item $\text{wt}$ - Hamming weight
	\item $\cat$ - string concatenation
	\item $com$ - string commitment function
	\item reductions - the operations in our protocols involve modular reduction by $q$.
	\item $\rexec S$ - choose uniformly at random from set $S$
  \item $\chiexec S$ - choose at random from set $S$ with distribution $\chi$ according 
  \item $S_n$ - set of permutations over $\{1,\ldots,n\}$
  \item $\triangleright$ - comment in pseudo code
  \item $\checkE$ - check if an equality holds
\end{itemize}

\section{Our LWE-based Scheme}
\subsection{Version with 2/3 soundness error}

\subsubsection{Key-Generation Algorithm}
\begin{algorithmic}[1]
\Procedure{KeyGen}{$n,m,q$}
\State $\matrix{A} \rexec \mathbb{F}_q^{n \times m}$, ~$\vector{s} \rexec \mathbb{F}_q^{m}$, ~$e  \chiexec  \mathbb{F}_q^{n}$.
\State $\vector{b} \gets \matrix{A}\vector{s}+\vector{e}$
\State $p \gets \text{wt}(\vector{e})$
\State $\triangleright$ $(\matrix{A},\vector{b},p)$ is the public key
\State $\triangleright$ $(\vector{s},\vector{e})$ is the private key
\State \textbf{return} $\{(\matrix{A},\vector{b},p),(\vector{s},\vector{e})\}$
\EndProcedure
\end{algorithmic}


\subsubsection{Identification Algorithm}
\begin{algorithmic}[1]
\Procedure{Identification}{$n,m,q$}
\State $\{(\matrix{A},\vector{b},p),(\vector{s},\vector{e})\} \gets \textsc{KeyGen}(n,m,q)$
\State $\triangleright$ \textbf{Prover}
\State $\vector{u} \rexec \mathbb{F}_q^{m}, \vector{\gamma} \rexec \mathbb{F}_{q}^{m}$ with $\gamma_i \neq 0, \forall i$
\State $\Sigma \rexec S_n$
\State $\triangleright$ Compute the commitments
\State $\vector{c_1} \gets com( \Pi_{\gamma, \Sigma}; \vector{r_1} )$ 
\State $\vector{c_2} \gets com( \Pi_{\gamma, \Sigma}(\matrix{A}(\vector{u}+\vector{s})); \vector{r_2} )$
\State $\vector{c_3} \gets com( \Pi_{\gamma, \Sigma}(\matrix{A}\vector{u}+\vector{b}); \vector{r_3} )$
\State $\triangleright$ Send the commitments to the Verifier
\State $\triangleright$ \textbf{Verifier}
\State $ch \rexec \{1,2,3\}$.
\State $\triangleright$ Send the challenge $ch$ to the Prover.
\State $\triangleright$ \textbf{Prover}
\State $\triangleright$ Open the commitments to the Verifier
\If{$ch=1$}
   \State send $\vector{r_1}, \vector{r_2}, \vector{u}+\vector{s}$ and $\Pi_{\gamma, \Sigma}$
\ElsIf{$ch=2$}
   \State send $\vector{r_2}, \vector{r_3}, \Pi_{\gamma, \Sigma}(\matrix{A}(\vector{u}+\vector{s}))$ and $\Pi_{\gamma, \Sigma}(\vector{e})$
\ElsIf{$ch=3$}
   \State send $\vector{r_1}, \vector{r_3}, \Pi_{\gamma, \Sigma}$ and $\vector{u}$
\EndIf
\State $\triangleright$ \textbf{Verifier}
\State $\triangleright$ Check the commitments
\If{$ch=1$} 
   \State check that $\vector{c_1}$ and $\vector{c_2}$ are correct.
\ElsIf{$ch=2$} 
   \State check that $\vector{c_2}$ and $\vector{c_3}$ are correct;
   \State check that $\text{wt}(\Pi_{\gamma, \Sigma}(\vector{e}))=p$.
\ElsIf{$ch=3$} 
   \State check that $\vector{c_1}$ and $\vector{c_3}$ are correct.
\EndIf
\If{all the checks were correct} 
   \State \textbf{return} ``success''
\Else 
   \State \textbf{return} ``failure''.
\EndIf
\EndProcedure
\end{algorithmic}

%
%

\subsection{Properties}
In this section we prove the properties of completeness, soundness and zero-knowledge of our identification scheme.

\paragraph{Completeness}

~\\ \textit{Proof}:
Knowing the secret values $(\vector{s},\vector{e})$, 
an honest Prover $P_h$ 
has all the information needed in order to compute the commitments and answer the challenges that enable the honest 
Verifier $V_h$ 
to ascertain the validity of the commitments and the parameters from which they were obtained. Both parties have access to the public parameters:
$\{\matrix{A},\vector{b},com\}$. Thus,\
\[
Pr(\text{Identification Algorithm}(P_h,V_h) = \text{``success''})=1
\]
\qed
 
\paragraph{Soundness}

~\\Now we show that a cheating Prover $P_c$ cannot deceive an honest Verifier $V_h$ with probability strictly greater than 2/3 without breaking the security assumptions upon which the Identification Algorithm is built.

\begin{thm} \label{theorem1}
If $V_h$ accepts a proof from $P_c$ with probability $ \geq (\frac{2}{3})^r + \epsilon$, with $\epsilon$ non-negligible, then there exists
a polynomial time probabilistic machine M which, with overwhelming probability,
either computes a valid secret pair $(\vector{s}, \vector{e})$ or finds a collision in the commitment scheme.
\end{thm}

Let us suppose, by absurd, that the cheating Prover $P_c$ is able to get accepted with probability $ \geq (\frac{2}{3})^r + \epsilon$, with $\epsilon$ non-negligible. Similarly to the proof technique applied by V\'{e}ron \cite{DBLP:journals/aaecc/Veron96}, we show that, either a collision has been found in the underlying commitment
scheme or we can find in polynomial time a set of commitments to which the cheating Prover $P_c$ is able to provide answers that enable the Verifier $V_h$ to re-compute them, retrieving the private secret $(\vector{s}, \vector{e})$, thus violating the hardness of the LWE problem.

Let us use subscripts $\{a,b,c\}$ to denote the answers given to challenges $\{1,2,3\}$. 

If one is able to find collisions in the commitment scheme, then it is possible to come up with different answers that reproduce the same set $\{\vector{c_1}, \vector{c_2}, \vector{c_3}\}$.

In the other case, prover $P_c$ is supposed to provide answers so that the equalities below are satisfied:
\[
\Pi_{\vector{\gamma_a}, \Sigma_a}=\Pi_{\vector{\gamma_c}, \Sigma_c}, \text{~for~} c_1;
\]
\[
\Pi_{\vector{\gamma_b}, \Sigma_b}(\matrix{A}(\vector{u_b}+\vector{s_b}))=\Pi_{\vector{\gamma_a}, \Sigma_a}(\matrix{A}(\vector{u_a}+\vector{s_a})) \text{~for~} c_2;
\]
\[
\Pi_{\vector{\gamma_b}, \Sigma_b}(\matrix{A}(\vector{u_b}+\vector{s_b}))+\Pi_{\vector{\gamma_b}, \Sigma_b}(\vector{e_b})=\Pi_{\vector{\gamma_c}, \Sigma_c}(\matrix{A}\vector{u_c}+\vector{b}) \text{~for~} c_3.
\]

From these, one can derive the secret parameters as follows:
\[
\vector{s} \leftarrow (\vector{u_a} + \vector{s_a})-\vector{u_c};
\]
\[
\vector{e} \leftarrow \Pi_{\vector{\gamma_c}, \Sigma_c}^{-1}(\Pi_{\vector{\gamma_b}, \Sigma_b}(\vector{e_b})).
\]

Let now $RA$ denote the random tape from which Prover $P_c$ obtains the elements necessary to compute the commitments and answers, and let $Q$ be the set $\{0,1,2\}$ from which the Verifier $V_h$ obtains his/her challenges. Denote by $(f,g) \in (RA \times Q)$ a pair chosen uniformly at random from these sets. We consider such a pair to be valid, that is $(f,g) \in \text{\emph{Valid}}$, when it leads to a result equal to ``success'' after $r$ rounds of execution. 
We have assumed in the beginning of this proof that the cardinalities of the sets involved are such that
\[
 \frac{\text{card}(\text{\emph{Valid}})}{\text{card}(RA \times Q)^r} \ge \left(\frac{2}{3}\right)^r + \epsilon.
\]

One can reason with the pigeonhole principle to characterize the ability of the prover $P_c$ to answer to strictly more than two challenges for any set of commitments. It can be modeled as follows. Let $\Omega_r \in {RA}^r$
such that 
\begin{itemize}
	\item When $f \in \Omega_r$, we have $2^r+1 \leq \text{card}(g) \leq 3^r$, with $(f,g) \in \text{\emph{Valid}}$
	\item When $f \in RA^r\backslash\Omega_r$, we have $0 \leq \text{card}(g) \leq 2^r$, with $(f,g) \in \text{\emph{Valid}}$
\end{itemize}

Hence,
\[
\frac{\text{card}(\text{\emph{Valid}})}{\text{card}((RA \times Q)^r)} \leq \frac{\text{card}(\Omega_r)}{\text{card}(RA^r)} + \left( \frac{2}{3}\right)^r.
\]

From the assumption in this proof,
\[
\frac{\text{card}(\Omega_r)}{\text{card}(RA^r)} \geq \epsilon.
\]

By resetting the cheating Prover $P_c$ an average number of times equal to $1/\epsilon$, it is possible to find an execution instances such that he has to answer to three different challenges for the same set of commitments. With that, we would have obtained in polynomial time a solution for any LWE instance.
\qed

\paragraph{Zero-Knowledge}
~\\
Let us build a simulator $S$ that mimics the communication tape between the Prover $P$ and the Verifier $V$, and then show that it cannot be statistically
distinguished from a real tape. Let us call $(P,V)$ the real communication tape, and $(P^{\prime},V^{\prime})$ its simulation. We assume that there is access to the verifier $V$ as a black box: it is fed with the public parameters and gives as response a challenge.

\begin{algorithmic}[1]
\Procedure{Simulator}{$A,b,p,n,m,q,r$}
	     \State $ch^{\prime} \rexec \{1,2,3\}$ \Comment predict the challenge.
       \State $\vector{\gamma} \rexec \mathbb{F}_{q}^{m}$ with $\gamma_i \neq 0, \forall i \in \{1,\ldots,m\}$
       \State $\Sigma \rexec S_m$, ~$\vector{u}^{\prime} \rexec \mathbb{F}_q^m$, ~$\vector{e}^{\prime} \chiexec \mathbb{F}_q^n$, with $\text{wt}(\vector{e}^{\prime})=p$
	     \State $\vector{r_1} \rexec \{0,1\}^n$,~$\vector{r_2} \rexec \{0,1\}^n$,~$\vector{r_3} \rexec \{0,1\}^n$

	     \If{$ch^{\prime}=1$} \Comment{prepare $\{\vector{c_1}, \vector{c_2}\}$}
	     \State Solve $\vector{y}$ satisfying $\matrix{A}\vector{y} = \vector{b}-\vector{e}^{\prime}$
	     \State Compute $\vector{c_1} \leftarrow com( \Pi_{\vector{\gamma}, \Sigma}; \vector{r_1} )$
	     \State Compute $\vector{c_2} \leftarrow com( \Pi_{\vector{\gamma}, \Sigma}(A\vector{y}; \vector{r_2} ))$
	     \State $\vector{c_3} \rexec \text{~Image of~}com$

	\ElsIf{$ch^{\prime}=2$} \Comment{prepare $\{\vector{c_2}, \vector{c_3}\}$}
	     \State Solve $\vector{s}^{\prime}$ satisfying $\matrix{A}\vector{s}^{\prime} = \vector{b}-\vector{e}^{\prime}$
       \State $\vector{c_2} \gets com( \Pi_{\vector{\gamma}, \Sigma}(A(\vector{u}^{\prime}+\vector{s}^{\prime})); \vector{r_2} )$
       \State $\vector{c_3} \gets com( \Pi_{\gamma, \Sigma}(A\vector{u}^{\prime}+\vector{b}); \vector{r_3} )$
	     \State $\vector{c_1} \rexec \text{~Image of~}com$

	\Else \Comment{prepare $\{\vector{c_1}, \vector{c_3}\}$}
	     \State $\vector{c_1} \gets com( \Pi_{\vector{\gamma}, \Sigma}; \vector{r_1} )$
       \State $\vector{c_3} \gets com( \Pi_{\vector{\gamma}, \Sigma}(A\vector{u}^{\prime}+\vector{b}); \vector{r_3} )$
       \State $\vector{c_2} \rexec \text{~Image of~}com$
   \EndIf
     
   \State $ch \gets V^{\prime}(\vector{c_1},\vector{c_2},\vector{c_3})$.
   \If{$ch$ and $ch^{\prime}$ are different}
      \State rewind $V^{\prime}$ and goto step 2
   \Else
   \State Open the commitments and save the messages.
   \State $r \gets r-1 $
   \EndIf
   \If{$r>0$} 
        \State Go to step 2.
   \EndIf        
\EndProcedure
\end{algorithmic}

The values chosen by the simulator in order to compute the commitments follow the same distribution as that from a real execution. The statistically hiding property of the commitment scheme $com$ conceals the fact that some of the commitments were just taken as random values, instead of actually computed via the application of $com$ to some set of parameters. Therefore, the transcript of the simulation above is statistically indistinguishable from what would have been obtained from a real execution of the protocol, proving that it has the property of statistical zero-knowledge.
\qed

%
%


\subsection{Version with 1/2 soundness error}
Cayrel, V\'{e}ron and El Yousfi \cite{DBLP:conf/sacrypt/CayrelVA10} proposed an identification scheme with soundness error approximately $1/2$, using the hardness of syndrome decoding over $q$-ary code as security assumption.

Here, we revisit their construction, adapting the core protocol to work over LWE. We also suggest the use of a lattice-based string commitment scheme \cite{DBLP:conf/crypto/ApplebaumCPS09}, aiming at applying a single security assumption: the hardness of LWE. These changes allowed us to obtain aa scheme whose security is based on a problem for which there is a quantum reduction from worst-cases of hard lattice problems \cite{DBLP:conf/coco/Regev10}. In order to speed up the operations involving multiplications with matrices and vectors, as well as reducing memory footprint, we adopt the use of rings. \cite{DBLP:conf/eurocrypt/LyubashevskyPR10}

\subsubsection{Key Generation Algorithm}

\begin{algorithmic}[1]
\Procedure{KeyGen}{$n,m,q$}
\State $\matrix{A} \rexec \mathbb{F}_q^{n \times m}$, ~$\vector{s} \rexec \mathbb{F}_q^{m}$, ~$e  \chiexec  \mathbb{F}_q^{n}$
\State $\vector{b} \gets \matrix{A}\vector{s}+\vector{e}$
\State Compute $\matrix{A}^{\bot}$ such that $\matrix{A}\matrix{A}^{\bot}=0$
\State $\vector{y} \gets \matrix{A}^{\bot}\vector{e}$; $p \gets \text{wt}(\vector{e})$
\State $\triangleright$ $(\matrix{A},\matrix{A}^{\bot},\vector{y},\vector{b},p)$ is the public key
\State $\triangleright$ $(\vector{s},\vector{e})$ is the private key
\State \textbf{return} $\{(\matrix{A},\matrix{A}^{\bot},\vector{y},\vector{b},p),(\vector{s},\vector{e})\}$
\EndProcedure
\end{algorithmic}

Unless stated otherwise, the random choices assume that the distribution is uniform. In order to hide the private parameters involved in the messages
exchanged between the Prover and the Verifier, we use three mechanisms:
\begin{itemize}
\item[i] a computationally binding and statistically hiding commitment scheme, denoted by $com$;
\item[ii] a weight-preserving transformation $\Pi_{\gamma, \Sigma}$, as defined in \ref{wtprestrans};
\item[iii] a blinding sum with a random factor uniformly chosen.
\end{itemize}

\subsubsection{Identification Algorithm}
\begin{algorithmic}[1]
\Procedure{Identification}{$n,m,q$}
\State $\{(\matrix{A},\matrix{A}^{\bot},\vector{y},\vector{b},p),(\vector{s},\vector{e})\} \gets \textsc{KeyGen}(n,m,q)$
\State $\triangleright$ \textbf{Prover}
\State $\vector{u} \rexec \mathbb{F}_q^{n}$
\State $\vector{\gamma} \rexec \mathbb{F}_{q}^{n}$  with $\gamma_i \neq 0, \forall i$
\State $\Sigma \rexec S_n$ 
\State $\triangleright$ Compute the commitments
\State $\vector{c_1} \leftarrow com( \vector{\gamma}  \cat  \Sigma  \cat \matrix{A}^{\bot}\vector{u} ; \vector{r_1} )$ 
\State $\vector{c_2} \leftarrow com( \Pi_{\vector{\gamma}, \Sigma}(\vector{u}) \cat \Pi_{\vector{\gamma}, \Sigma}(e); \vector{r_2} )$
\State $\triangleright$ Send the commitments to the Verifier
\State $\triangleright$ \textbf{Verifier}
\State $\alpha \rexec \mathbb{Z}_q$.
\State Send  $\alpha$ to the Prover.
\State $\triangleright$ \textbf{Prover}
\State Respond with $\vector{\beta} \gets \Pi_{\vector{\gamma}, \Sigma}(\vector{u}+\alpha \vector{e})$
\State $\triangleright$ \textbf{Verifier}
\State Send a challenge $ch \in \{1,2\}$
\State $\triangleright$ \textbf{Prover}
\State $\triangleright$ Open the corresponding commitment
\If{$ch=1$}
	\State respond with $\vector{r_1}, \vector{\gamma}, \Sigma$
\ElsIf{$ch=2$} 
        \State respond with $\vector{r_2}, \Pi_{\vector{\gamma}, \Sigma}(\vector{e})$
\EndIf
\State $\triangleright$ \textbf{Verifier}
\State $\triangleright$ Check the commitments
\If{$ch=1$}
	\State $\vector{c_1} \checkE com(\Sigma \cat \vector{\gamma} \cat \matrix{A}^{\bot}\Pi^{-1}_{\gamma,\Sigma}(\vector{\beta}) - \alpha \vector{y}; \vector{r_1})$
\ElsIf{$ch=2$} 
\State $\vector{c_2} \checkE com(\vector{\beta}-\alpha\Pi_{\vector{\gamma},\Sigma}(\vector{e}) \cat \Pi_{\gamma,\Sigma}(\vector{e}); \vector{r_2})$ 
\State $\text{wt}(\Pi_{\gamma,\Sigma}(\vector{e})) \checkE p$
\EndIf
\If{all the checks were correct} 
   \State \textbf{return} ``success''
\Else 
   \State \textbf{return} ``failure''.
\EndIf
\EndProcedure
\end{algorithmic}

\subsection{Properties}
In this section we prove the properties of completeness, soundness and zero-knowledge of our identification scheme.

\paragraph{Completeness}

~\\ \textit{Proof}:
The knowledge of the secret values $(\vector{s},\vector{e})$ enables an honest Prover $P_h$ to compute the commitments and answer any  challenge that the honest Verifier $V_h$ may pose. Both parties have access to the public parameters $\{\matrix{A}^{\bot},\vector{y},com\}$. Thus,\
\[
Pr(\text{Identification Algorithm}(P_h,V_h) = \text{``success''})=1
\]
\qed
 
\paragraph{Soundness}

~\\The structure of the protocol is essentially that from Cayrel et al. \cite{DBLP:conf/sacrypt/CayrelVA10}, except for the way the commitments are computed and the underlying hard problem. The reasoning about the relative size of the sample spaces from which the random choices are made follow a similar line. The main difference rests in the way the secret keys are extracted from the commitments once they are opened by the Prover, as shown below.

\begin{thm} \label{theorem2}
If $V_h$ accepts a proof from $P_c$ with probability $ \geq (\frac{q+1}{2q})^r + \epsilon$, with $\epsilon$ non-negligible, then there exists
a polynomial time probabilistic machine M which, with overwhelming probability,
either computes the secret value $e$ or finds a collision in the commitment scheme. 
\end{thm}

We use subscript $a$ to denote the values revealed upon reception of challenge equal to 1, and subscript $b$ for challenge equal to 2. Then
 
\[
\Pi_{\vector{\gamma_a}, {\Sigma}_a}=\Pi_{\vector{\gamma_b}, \Sigma_b}, \text{~for~} c_1,
\]
\[
\Pi_{\vector{\gamma_a}, \Sigma_a}(\vector{e_a})=\Pi_{\vector{\gamma_b}, \Sigma_b}(\vector{e_b}) \text{~for~} c_2.
\]

Given that $\{\vector{\gamma_a}, \Sigma_a\}$ are published due to the challenge equal to 1, and $\Pi_{\vector{\gamma_b}, \Sigma_b}(\vector{e_b})$ is published due to the challenge equal to 2, such information can be used to derive the secret parameter $e$ as follows:
\[
\vector{e} \leftarrow \Pi_{\vector{\gamma_a}, \Sigma_a}^{-1} (\Pi_{\vector{\gamma_b}, \Sigma_b}(\vector{e_b})).
\]

From the assumption made in this proof,
\[
\frac{\text{card}(\Omega_r)}{\text{card}(RA^r)} \geq \epsilon.
\]

By resetting the cheating Prover $P_c$ an average number of times equal to $1/\epsilon$, it is possible to find an execution instances such that he has to answer to three different challenges for the same set of commitments. With that, we would have obtained in polynomial time a solution for any LWE instance.
\qed

\paragraph{Zero-Knowledge}
~\\
Let us build a simulator $S$ that mimics the communication between the Prover $P$ and the Verifier $V$, and then show that it cannot be statistically
distinguished from a real tape. Let us call $(P,V)$ the real communication tape, and $(P^{\prime},V^{\prime})$ its simulation. We assume that there is access to the verifier $V$ as a black box: it is fed with the public parameters and gives as response a challenge.

The simulator is built as follows, using oracle access to a verifier $V$.
\begin{algorithmic}[1]
\Procedure{Simulator}{$A,A^{\bot},b,y,p,n,m,q,r$}
   \State $\vector{\gamma} \rexec \mathbb{F}_{q}^{m}$ with $\gamma_i \neq 0, \forall i \in \{1,\ldots,m\}$
   \State $\Sigma \rexec S_m$, ~$ch^{\prime} \rexec \{1,2\}$, ~${\vector{u}^\prime} \rexec \mathbb{F}_q^{m}$
   \State $\vector{r_1} \rexec \{0,1\}^n$, ~$\vector{r_2} \rexec \{0,1\}^n$
   \If{$ch^{\prime}=1$}
	   \State Solve $\vector{e}^{\prime}$ for $\matrix{A}^{\bot} \vector{e}^{\prime}= \vector{b}$
	   \State $\vector{c_1} \gets com( \vector{\gamma}  \cat  \Sigma  \cat \matrix{A}^{\bot}\vector{u}^{\prime}; \vector{r_1})$ 
	   \State $\vector{c_2} \rexec \text{~Image of~}com$
   \Else
	   \State $\vector{e}^{\prime} \rexec \mathbb{F}_{q}^{m}$ with  $\text{wt}(\vector{e}^{\prime})=p$
	   \State $\vector{c_2} \gets com( \Pi_{\vector{\gamma}, \Sigma}(\vector{u}^{\prime}) \cat \Pi_{\vector{\gamma}, \Sigma}(\vector{e}^{\prime}); \vector{r_2} )$
	   \State $\vector{c_1} \rexec \text{~Image of~}com$
   \EndIf
   \State $\alpha \gets V^{\prime}(\vector{c_1},\vector{c_2})$
   \State $\vector{\beta} \gets \Pi_{\vector{\gamma}, \Sigma}(\vector{u}^{\prime}+\alpha \vector{e}^{\prime})$ 
   \State  $ch^{\prime} \gets V^{\prime}(\vector{c_1},\vector{c_2},\vector{\beta})$
   \If{$ch$ and $ch^{\prime}$~are different} 
      \State rewind $V^{\prime}$~and go to step 2
   \Else
      \State Open the commitments and save the messages.
      \State $r \gets r-1 $
   \EndIf
   \If{$r>0$} 
        \State Go to step 2.
   \EndIf        
\EndProcedure
\end{algorithmic}

The values chosen by the simulator in order to compute the commitments follow the same distribution as that from a real execution. The statistically hiding property of the commitment scheme $com$ conceals the fact that some of the commitments were just taken as random values. Therefore, the transcript of the simulation above is statistically indistinguishable from what would have been obtained from a real execution of the protocol, proving that it has the property of statistical zero-knowledge.
\qed


\subsection{Security and Performance}
\subsubsection{Overall Soundness Error}
There are two security aspects to be taken into account regarding the scheme: the hardness of the underlying LWE problem and the overall soundness error.
The first factor is linked with the values of $\{n,m,q\}$ and can be determined using the best known algorithm for solving LWE \cite{DBLP:conf/ctrsa/LindnerP11}. The analysis of Lindner and Peikert address an encryption scheme but we believe they can be adapted to our setting. The second is related to the desired upper bound $L$ for the probability of success for an impersonation after $r$ rounds of protocol execution. It has a direct impact in the communication costs, given that the following condition must be met: 
\begin{itemize}
\item $(2/3)^r \leq L$~for the first scheme;
\item $ \left( \frac{q+1}{2q} \right)^r \leq L$~for the second scheme.
\end{itemize}

Our system is secure under the Serial Active Model.


\subsubsection{Communication Costs}
Let us calculate the average communication costs for this identification scheme. Whenever a random vector is supposed to be exchanged between the Prover and the Verifier, we send the corresponding seed from which the element can be obtained, assuming that both parties agree upon the use of a pseudo-random generator. The definition of the isometry $\Pi_{\vector{\gamma,\Sigma}}$ takes two seeds (one for the vector $\vector{\gamma}$ and other for the permutation $\Sigma$). The number of bits of a given element is returned by the application of $\left|seed\right|$.  Therefore, the payload breakdown per round of execution can be seen as follows:
\begin{itemize}
 \item Commitments: $3\left|com\right|$
 \item Challenge: $\left\lceil \log_2{\text{max}(ch)} \right\rceil$
 \item Answer (avergage): $\frac{10}{3}\left|seed\right|+\frac{2}{3}(m+n)\left\lceil \log_2{q}\right\rceil$
\end{itemize}

Similarly to the procedure followed with the 2/3 soundness error scheme, we have for the 1/2 soundness error version the following breakdown for the communication costs per round:
\begin{itemize}
 \item Commitments: $2\left|com\right| + n\left\lceil \log_2{q}\right\rceil$
 \item Challenges: $\left\lceil \log_2{\text{max}(ch)}  \right\rceil + \left\lceil \log_2{q}\right\rceil$
 \item Answer (average): $2\left|seed\right|+\frac{n}{2}\left\lceil \log_2{q}\right\rceil$
\end{itemize}


\section{Conclusion}
We have shown in this paper an adaptation for lattices of two zero-knowledge identification schemes originally designed with codes. Using the hardness of LWE as security assumption and a set of suitable parameters, we obtained a construction with worst-case connection with hard lattice problems. Through the use of ring-LWE constructions, the memory footprint is taken to levels similar to what could be obtained with ideal-SIS schemes and operations involving multiplication with vectors are more efficiently performed via FFT. Besides, the adaptations preserved much of the structure of the original protocols. From a theoretical angle, this points towards a possible unification of cryptographic schemes based on codes and lattices.


%
%



%
\bibliographystyle{plain}
\bibliography{bare_conf}

\end{document}